\pdfoutput=1

\documentclass[%
 a4paper,
 reprint,
 onecolumn,
superscriptaddress,
 amsmath,amssymb,
apl,
pra,
]{revtex4-2}

\PassOptionsToPackage{round}{natbib}
\setcitestyle{authoryear}
\usepackage[T1]{fontenc} 
\usepackage[utf8]{inputenc}
\usepackage{lmodern} 
\usepackage[english]{babel}
\usepackage[babel=true]{microtype}
\usepackage{graphicx}
\usepackage[svgnames,table]{xcolor}
\usepackage{dcolumn}
\usepackage{bm}
\usepackage{float}
\usepackage{array, booktabs}
\usepackage[round,sort&compress]{natbib}

\graphicspath{{figures/}}

\begin{document}
\title{Assimilation of atmospheric infrasound data to constrain tropospheric and stratospheric winds}

\author{Javier Amezcua}
 \email{j.amezcuaespinosa@reading.ac.uk}
 \affiliation{UK National Centre for Earth Observation, University of Reading, Reading, UK}
\author{Sven Peter N{\"a}sholm}%
\email{peter@norsar.no}
\affiliation{%
NORSAR, Kjeller, Norway}

\author{Erik M{\aa}rten Blixt}
\email{Present address: Blixt Geo AS, Oslo, Norway}
\affiliation{%
NORSAR, Kjeller, Norway}

\author{Andrew J. Charlton-Perez}
\affiliation{%
Department of Meteorology, University of Reading, Reading, UK}

\date{April 16, 2020} 

\begin{abstract}
This data assimilation study exploits infrasound from explosions to probe an atmospheric wind component from the ground up to stratospheric altitudes. %
Planned explosions of old ammunition in Finland generate transient infrasound waves that travel through the atmosphere. These waves are partially reflected back towards the ground from stratospheric levels, and are detected at a receiver station located in northern Norway at 178\,km almost due North from the explosion site. %
The difference between the true horizontal direction towards the source and the backazimuth direction (the horizontal direction of arrival) of the incoming infrasound wave-fronts, in combination with the pulse propagation time, are exploited to provide an estimate of the average cross-wind component in the penetrated atmosphere. %
We perform offline assimilation experiments with an ensemble Kalman filter and these observations, using the ERA5 ensemble reanalysis atmospheric product as background (prior) for the wind at different vertical levels. %
We demonstrate that information from both sources can be combined to obtain analysis (posterior) estimates of cross-winds at different vertical levels of the atmospheric slice between the explosion site and the recording station. 
The assimilation makes greatest impact at the $12-60$\,km levels, with some changes with respect to the prior of the order of $0.1-1.0$\,m/s, which is a magnitude larger than the typical standard deviation of the ERA5 background. The reduction of background variance in the higher levels often reached $2-5\%$. 
This is the first published study demonstrating techniques to implement assimilation of infrasound data into atmospheric models. It paves the way for further exploration in the use of infrasound  observations -- especially natural and continuous sources -- to probe the middle atmospheric dynamics and to assimilate these data into atmospheric model products.  
\par
\smallskip
\noindent\emph{Manuscript submitted to the Quarterly Journal of the Royal Meteorological Society.}
\par
\smallskip
\noindent\emph{The current document is an e-print typeset by the authors.}
\end{abstract}


\maketitle

\newcommand{\dd}{\ensuremath{\,\text{d}}}
\newcommand{\lat}{\ensuremath{\text{lat}}}
\newcommand{\lon}{\ensuremath{\text{lon}}}
\newcommand{\wc}{\ensuremath{\bar W_c}}
\newcommand{\celerity}{\ensuremath{\upsilon}}

\section{\label{sec:background}Introduction}
Despite much recent attention to extra-tropical stratospheric dynamics and their connection to the troposphere, the amount of observational data in the stratosphere available to numerical weather prediction centres remains limited. A better representation of the stratospheric dynamics and the stratosphere-troposphere coupling in models has the potential to enhance tropospheric weather forecasts, in particular, on sub-seasonal timescales \citep{Baldwin2003, polavarapu2005some, Charlton2007,mitchell2013influence, kidston2015stratospheric,  karpechkoWMO2016, haase2018importance,  blanc2018toward, Pedatella2018, taguchi2018seasonal, kawatani2019effects}. %
Moreover, the lid of several atmospheric model products has been raised into the mesosphere \citep{polavarapu2005some} 
and it has been demonstrated that this can improve numerical weather and climate models \citep{orsolini2011potential, kidston2015stratospheric, charlton2013lack}. %
But the full potential of high-top models can only be unlocked if middle atmospheric winds are better represented 
\citep{baker2014lidar,lee2019potential,korhonen2019strat-discussion}. %
Hence, it is timely to explore novel datasets and assimilation approaches that can constrain the upper stratospheric dynamics in atmospheric model products. %

Infrasound waves are acoustic waves at frequencies below the human hearing limit (typically around 20\,Hz). %
These waves can be generated by natural sources, such as volcanoes, earthquakes and ocean swell, but also by human sources, such as mining and explosions (see, e.g. \cite{Pichon:2018}). %
These waves propagate through the atmosphere and can be recorded by ground-based stations. The wave frequencies of greatest interest for atmospheric characterisation are typically of the order of 1\,Hz. %
The time and form of the received signals provide temperature and wind related information about the atmosphere the waves traverse. %
Infrasound waves may travel from sources on the surface of the Earth, reach a maximum altitude where they are partly or fully reflected or refracted, and then reach back to the surface to be detected by a receiver. %
Effectively, they probe a slab of the atmosphere in a tomographic fashion since the time it takes for these waves to complete their path is affected by the characteristics of the atmosphere they pass through: in particular, the wind velocity and temperature, but also attenuation-related properties like density and relative humidity. %
Hence, spatio-temporally integrated information carried by the propagating infrasound waves can be utilised to reconstruct or constrain atmospheric variables. Sound waves are already exploited in other tomographic and imaging problems. For instance, in underwater acoustics, temperature profiles \citep{dzieciuch2013structure} and seafloor bathymetries \citep{Wolfl_etal_2019} are mapped using sound waves. Probabilistic infrasound propagation has been studied in \cite{Smets2015}, where measured infrasound wavefront parameters for one year of infrasound explosions were compared to raytracing simulations using the ensemble atmospheric wind and temperature fields of the ECMWF ensemble data assimilation system of perturbed analyses \citep{buizza1999stochastic}.

The current study follows directly from a recent paper by \citet{blixt2019estimating}, which used the same dataset to demonstrate that atmospheric cross-winds can be estimated directly from infrasound data using propagation time and backazimuth deviation observations, and interpreted these results in the context of ERA-Interim reanalysis winds. %
There is a physical effect which is the basis of this work: when a steady cross-wind acts on a propagating acoustical plane wave, a bending of the wave-front is introduced. This creates a deviation in the apparent backazimuth direction of infrasound wave-fronts impinging on ground-based sensor array stations. %
We use this physical effect to assess the dynamical evolution of the stratosphere during several events, as sampled by the infrasound waves on their paths between Finland and a ground-based station in Northern Norway. %
The array signal processing algorithms exploit infrasound signals recorded on a set of 25 sensors distributed on the ground within a 3\,km wide aperture (see \citet{blixt2019estimating}, Figure 1). %

Data assimilation (DA, e.g. \cite{Asch_et_al2016, Kalnay03atmosphericmodeling}) is a discipline which aims to combine different imperfect and incomplete sources of information to produce a better estimate of a variable of interest. In particular, it takes into account the uncertainty of the information sources. %
The most ambitious approach obtains and updates descriptions of a system using probability distribution functions (pdf's) by application of Bayes' theorem. %
In practice, however, sample estimators like mean, covariance and mode of the distributions often suffice. %
In particular, the Kalman filter \citep{kalman1960,Kalman61newresults} and its ensemble implementation \citep{Evensen1994,Burgers_etal1998, Tippett_et_al203} assume Gaussian statistics in the sources of errors, as well as none or small deviations from linearity in the evolution and observation processes. The filter operates with the first two statistical moments of a distribution. %
An advantageous feature of the Kalman filter is that it can assimilate an integrated observation variable (in our case an average wind component resulting from vertical integration along the path of propagation) and translate this into increments at different vertical levels. This proved useful, for instance, in the assimilation of radiance satellite observations \citep{Lei_etal2018}. %
A discussion on the prospects of assimilating atmospheric infrasound data into numerical weather prediction models can be found in \citet{Assink2019advances}.

There are two main objectives of this study. The first is to develop a framework which allows for assimilation of tropospheric and stratospheric wind information based on atmospheric infrasound data. The second is to provide a first demonstration and proof-of-concept with an offline (i.e.\ no cycling involved) infrasound DA experiment using the developed framework, exploiting a dataset which is already well-characterised in previous works.

We generate an estimate of the averaged cross-wind component along the relevant track from the explosion site in Finland to the station in Northern Norway, as well as an associated measure of uncertainty. We apply the deterministic ensemble Kalman Filter (DEnKF) as described in \cite{sakov_oke_denkf2008}. We select this approach because it allows for model-space localisation, as opposed to observation-space localisation which is not feasible for integrated quantities \citep{Lei_etal2018}. Some specifics of this method are outlined in Appendix \ref{sec:appendix}.

The remainder of this paper is organised as follows: Section \ref{sec:elements} explains the system setup, detailing the way observations are related to the state variables of the system under different degrees of simplification from the most general problem to the case considered in the current work. %
In Section \ref{sec:synthetic}, we perform synthetic-data experiments under ideal conditions with an infinite ensemble size, and with different vertical weights in the observation operator. %
These experiments verify the offline DA process in a controlled setting. %
Section \ref{sec:real_data} presents the real-data assimilation experiments using infrasound from 18 years of explosions. %
In Section \ref{sec:conclusions} we conclude the study, discuss its limitations, and provide ideas and suggestions for future work.

\section{\label{sec:elements}Cross-wind effects on the propagation and arrival of infrasound wavefronts}

Let us explore the effect of a cross-wind on the propagation of infrasound waves. Recall the basic principle: a background wind field affects the propagation of infrasound waves; specifically, a cross-wind can bend the wave front. Infrasound waves, however, do not modify the background wind field. 

\subsection{Propagation within a plane}
\begin{figure}[H]
   \centering
       \includegraphics[width=0.75\textwidth]{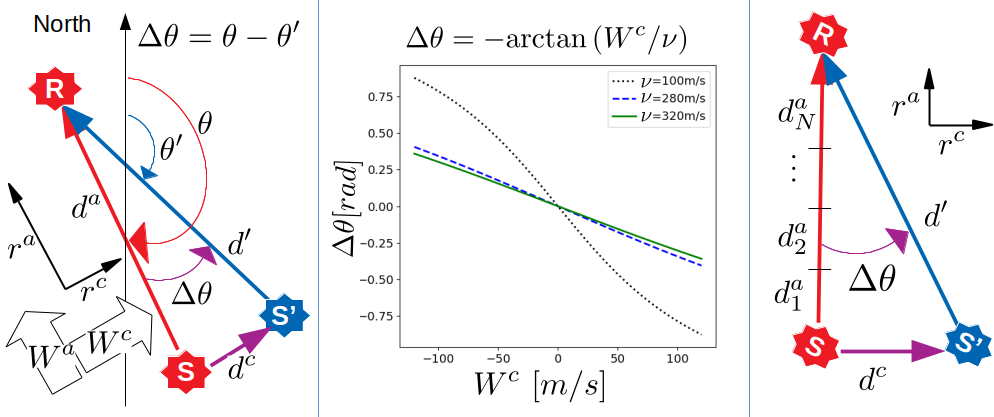}
   \caption{%
   Effect of cross-wind in the horizontal propagation of infrasound waves in a horizontal plane. The left panel shows the shift in backazimuth angle between the real source (S) and receiver (R) and the apparent source (S') and the receiver. The middle panel shows the relationship between change in backazimuth angle and the cross-wind for different celerity values. The right panel shows the same situation as the left panel, but after spatial discretisation. }
\label{fig:v2D}
\end{figure}

First, we discuss horizontal wave propagation only. We illustrate the situation in the left panel of Figure \ref{fig:v2D}. Consider a plane with two horizontal directions denoted $r^a$ and $r^c$; the sub-indices $a$ and $c$ denote \emph{along-track} and \emph{cross} respectively, and refer to the wind direction with respect to the propagation of the infrasound wave. The infrasound source is the red star in the bottom, labelled $S$, while the receiver is the red star in the top, labelled $R$. The straight red line connecting the two points has length $d^a$. The backazimuth $\theta$ is the angle of this line measured with respect to the North. Now consider a constant (for now) wind $W^c$ blowing perpendicular to the line $d^a$. The effect of $W^c$ is to create a change in the apparent backazimuth direction of infrasound wave fronts when they arrive to $R$ \citep{Diamond1964}. The received waves seem to come from an apparent source marked by the blue star $S'$, a distance $d'$ away from $R$, and with a modified backazimuth angle $\theta'$. The distance between the real and apparent sources (purple line) is $d^c$. The change in angle is denoted as:
\begin{equation}
 \Delta \theta = \theta'-\theta. \label{eq:Deltatheta} 
\end{equation}

\noindent
For positive cross-winds (as in the setup of the figure) the change in angle is negative. 

We can relate different elements of this system using the following considerations. The wave is emitted from $S$ and received at a time $T$ after the explosion. For infrasound waves propagating within atmospheric waveguides, \emph{celerity} $\celerity$ is defined as the ratio between the straight distance between source and receiver divided by travel time, i.e:
\begin{equation}
\celerity = \frac{d^a}{T}. \label{eq:defcel}
\end{equation}

\noindent 
The lines $d^a$ and $d^c$ are the two legs of a right-angled triangle. We can solve for $d^a$ from \eqref{eq:defcel}. For $d^c$ we simply have (since the cross-wind is constant):
\begin{equation}
d^c = W^c T. \label{eq:dcross}
\end{equation}

\noindent
Some trigonometry yields $\text{tan} \left( | \Delta \Theta | \right) = d^c / d^a$. Explicitly, this is:
\begin{equation}
 \Delta \theta = - \text{arctan} \left( \frac{W^c}{\celerity} \right). \label{eq:DTheta}
\end{equation}

\noindent
The negative sign comes from the direction $\Delta \theta$ that is defined in \eqref{eq:Deltatheta}. The middle panel of Figure \ref{fig:v2D} illustrates \eqref{eq:DTheta} for different values of cross wind (horizontal axis) and celerity (lines). Note that as long as $W^c \ll \celerity$ the function is close to linear. This is verified by the McLaurin expansion of the arctangent function:
\begin{equation}
\text{arctan} \left( \frac{W^{c}}{\celerity} \right) =  \frac{W^{c}}{\celerity} + \mathcal{O} \left( \frac{W^{c}}{\celerity} \right)^3. %
\label{eq:arctanMclaurin}
\end{equation}

So far, we have considered a constant cross-wind $W^c$. In the general case this speed can be a function of the position $r^a$ and time $t$, i.e. $w^c \left( r^a, t \right)$. Then \eqref{eq:dcross} becomes an integral:
\begin{equation}
d^c = \int_{0}^T w^c \left( r^a, t \right) dt. \label{eq:dc_int}
\end{equation}

\noindent
This computation is not easy in general. The position $r^a$ of the wave front depends on the infrasound speed of propagation along $d^a$, which is the sum of the sound speed (a function mainly of temperature) and the actual background wind along the direction of the propagation $w^a \left( r^a, t \right)$. 

\cite{blixt2019estimating} define an average cross-wind as:
\begin{equation}
W^c = \frac{1}{T} \int_{0}^T w^c\left( r^a, t \right) dt.
\end{equation}

\noindent
With this definition we can still use \eqref{eq:DTheta}, with $W^c$ being the average cross-wind velocity along $d^a$. It is actually more useful to convert this time integral into a spatial sum. We illustrate this process with the aid of the diagram in the right of figure \ref{fig:v2D}. We divide the line $d^a$ into $N$ segments. Each segment has length $d^a_n$, and it is clear that:
\begin{equation}
d^a = d^a_1 + d^a_2 + \cdots + d^a_N.
\end{equation}

\noindent
Similarly, the total travel time $T$ is the sum of the time spent in each segment:
\begin{equation}
T = T_1 + T_2 + \cdots + T_N.
\end{equation}

Consider also a constant background sound speed $\mathcal{C}$. Also, consider that the two wind components are constant per segment of $d^a$. In the $n^{th}$ segment $d^a_n$ we have: $\{ W^a_n, W^c_n \}$. The time for the infrasound wave to travel a segment $d^a_n$ is:
\begin{equation}
T_n = \frac{d^a_n} {\mathcal{C} + W^a_n}.
\end{equation}

\noindent
Following \eqref{eq:dcross}, the cross displacement in the $n^{th}$ segment is the product: $d^c_n = W^c_n T_n$. The total cross displacement is then the sum:
\begin{equation}
d^c = \sum_{n=1}^N {W^c_n \frac{d^a_n}{\mathcal{C}+W^a_n} }.  \label{eq:dc_patches}
\end{equation}

\noindent
Notice that by dividing the last expression by the total travel time $T$, we can define a weighted average cross wind speed as:
\begin{equation}
W^c = \sum_{n=1}^N \alpha_{n} W^c_n. \label{eq:wc_spatial}
\end{equation}

\noindent
with the weights:
\begin{equation}
\alpha_n = \frac{d^a_n}{T \left( \mathcal{C}+W^a_n \right)}. \label{eq:alpha_spatial}
\end{equation}

\noindent
Then, we can still use \eqref{eq:DTheta} to relate this weighted average to the change in backazimuth angle. Most importantly, we can estimate the average cross wind as a spatially weighted linear combination of cross winds. The weights derived in \eqref{eq:wc_spatial} follow from several simplifications. There are wave-tracing techniques that can model the trajectory of the propagating wave (see, e.g. \cite{hedlinWalker2013}) which can be used to determine adequate weights.  

\subsection{3D Propagation}
\begin{figure}[H]
   \centering
       \includegraphics[width=0.55\textwidth]{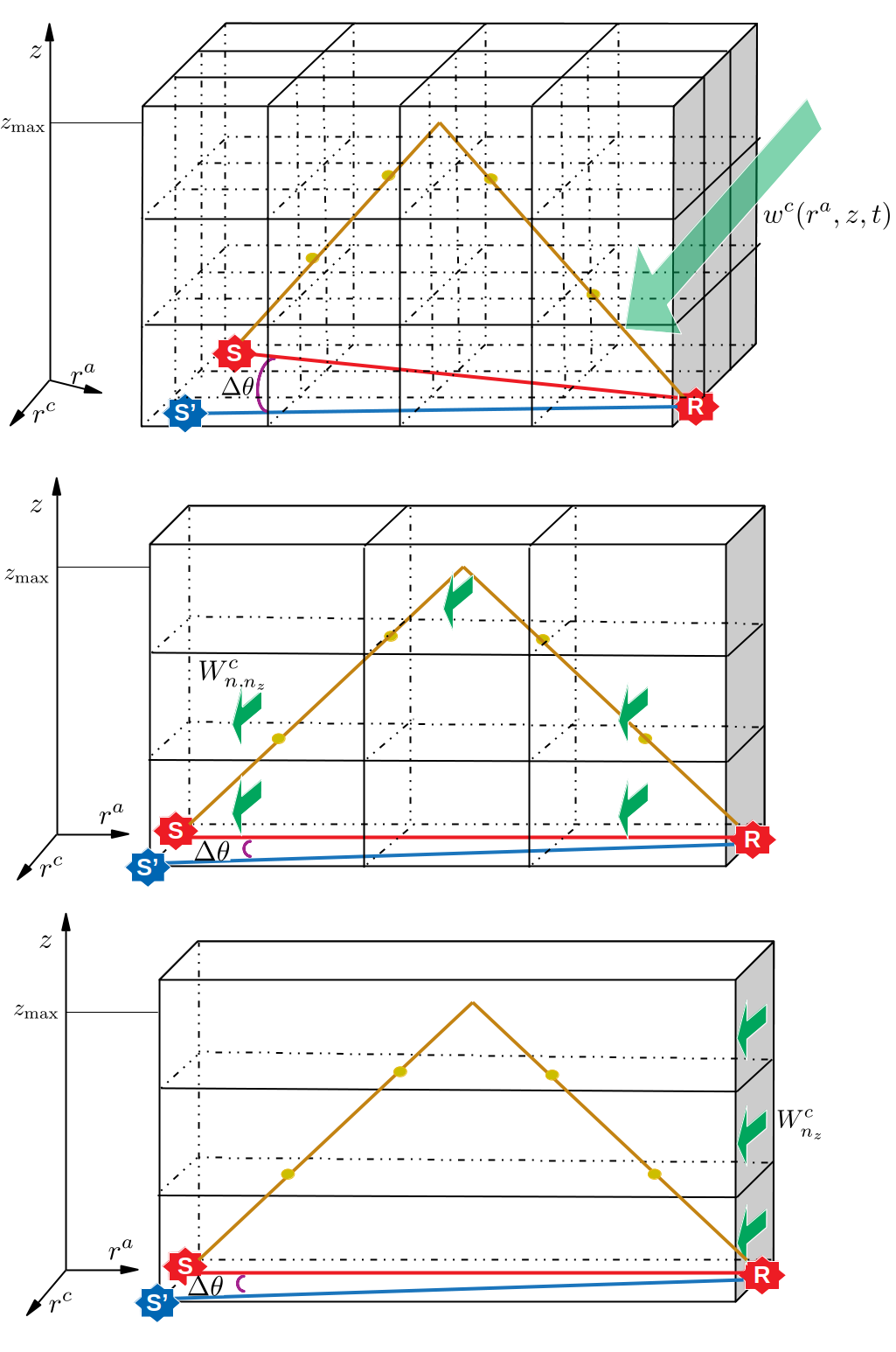}
   \caption{Depiction of an infrasound wave (yellow) travelling through an atmospheric volume. There are a source ($S$) and a receiver ($R$) at the surface. The wave travels vertically to a maximum altitude, where it is reflected, and it travels through a cross-wind field through all its trajectory. The top panel shows the original set-up with a native atmospheric grid. The middle panel shows the situation after reducing the problem to the along-track plane and discretising the (interpolated) cross-wind both in two directions. The bottom panel shows the problem after further simplifying by averaging the cross-winds on the along-track direction.} 
\label{fig:geometry}
\end{figure}

Having explained the basics, we now move to full 3D wave propagation, i.e. when the trajectory of the infrasound wave has a vertical component. This is depicted in Figure \ref{fig:geometry}. 
The top panel shows an atmospheric volume discretised to a model grid. Both the source ($S$) and receiver ($R$) are in the surface. In this case, the line $d^a$ is a segment of the great circle between $S$ and $R$, and it is not necessarily aligned with the grid. A simple example path of an infrasound wave is shown in yellow. The wave travels both in the $r^a$ and $z$ directions. The wave travels in the vertical to a given maximum altitude from where it returns down to ground (e.g. due to partial reflection as explained in \cite{blixt2019estimating}) and it is then detected at the receiver. As in the 2D case, the wave travels through a cross-wind field which leads to a change on backazimuth angle $\Delta \theta$ towards an apparent source $S'$. This cross-wind now also depends on altitude: $w^c(r^a,z,t)$. 

As before, the travel time $T$ is known, as well as the horizontal distance $d^a$, so we can still define celerity as in \eqref{eq:defcel}. The expression for $d^c$ is the same as \eqref{eq:dc_int}, but in this case the $w^c$ also depends on altitude: 
\begin{equation}
d^c = \int_{0}^T w^c \left( r^a,z, t \right) dt.
\label{eq:dc_intz}
\end{equation}

\noindent
Turning this time integral into a spatial sum is slightly more complicated. The process is illustrated in the middle panel of figure \ref{fig:geometry}. First, the situation is reduced to a 2D problem by creating a channel centred in the line $d^a$. The winds from the native grid are interpolated to provide the along and cross values in this new setup. As before, we divide the distance $d^a$ into $N$ different intervals, and this time the distance from $z=0$ to $z=z_{max}$ (which has to be determined) is divided into $N_z$ intervals. This creates a 2-dimensional DA grid where we consider the $\{n,n_z\}^{th}$ box to have a cross-wind $W^c_{n,n_z}$, which is obtained from the native grid via interpolation. It is important to notice that the wave does not go through all the boxes, but only a set of them, referred as $\Omega$ below. The displacement $d^c$ is affected only by the cross-wind in the boxes of this valid set. 

The expression for the average cross-wind, i.e. the equivalent to \eqref{eq:wc_spatial}, becomes a double sum:
\begin{equation}
W^c = \sum_{n_z=1}^{N_z} \sum_{n=1}^N \alpha_{n,n_z} W^c_{n,n_z}. \label{eq:wc_spatial_2ind}
\end{equation}

\noindent
The weight is zero for any box outside the set $\Omega$. For the boxes in the set $\Omega$ the weights are more complicated than in \eqref{eq:alpha_spatial}, since the (diagonal) length travelled by the wave in different boxes may be different, and the effective propagation speed includes both the along-track wind and vertical velocities. Therefore, one may rely on ray-tracing techniques to derive these weights. 

Since this is our first study, and the distance between source and receiver is relatively short ($178$\,km), we further simplify the problem as illustrated in the bottom panel of figure \ref{fig:geometry}. We do this by considering only $N=1$ interval along the propagation of the wave. For the rest of the work we consider the average cross-wind speed as a weighted sum over $N_z$ vertical levels:
\begin{equation}
W^c = \sum_{n_z=1}^{N_z} \alpha_{n_z} W^c_{n_z}. \label{eq:wc_spatial_vert}
\end{equation}

In DA terminology, our state variable is the vector of vertical (horizontally averaged) cross-winds $\mathbf{w}^c \in \mathcal{R}^{N_z}$:
\begin{equation}
\mathbf{w}^c = \begin{bmatrix} W^c_1 \\ W^c_2 \\ \vdots \\ W^c_{N_z}.
\end{bmatrix} \label{eq:wc_vector}
\end{equation}

\noindent
It is also useful to group the vertical weights in a vector as:
\begin{equation}
\boldsymbol{\alpha} = \begin{bmatrix} \alpha_1 \\ \alpha_2 \\ \vdots \\ \alpha_{N_z} \end{bmatrix}. \label{eq:alpha_vector}
\end{equation}

\noindent
In this case the sum \eqref{eq:wc_spatial_vert} is simply a vector product:
\begin{equation}
W^c = \boldsymbol{\alpha}^\mathbf{T} \mathbf{w}^c \label{eq:Wc_vec}
\end{equation}

\noindent
and \eqref{eq:DTheta}, in DA terms the observation equation, can simply be written as:
\begin{equation}
 \Delta \theta = - \text{arctan} \left( \frac{ \boldsymbol{\alpha}^\mathbf{T} \mathbf{w}^c }{\celerity} \right). \label{eq:obs_eq}
\end{equation}

\noindent
This is mapping that reduces dimensionality $\mathcal{R}^{N_z} \to R$.

\section{\label{sec:synthetic}Synthetic-data assimilation experiments}
This section describes basic synthetic DA experiments, before moving to the case of the assimilation of recorded infrasound data. %
Consider $N_z = 4$ vertical levels in a propagation volume. Let the cross-wind $\mathbf{w}^c \in \mathcal{R}^{4}$ be a Gaussian random variable with zero mean $\boldsymbol{\mu}^b = \mathbf{0}$ and covariance $\mathbf{B} \in \mathcal{R}^{4 \times 4}$. %
We apply the DEnKF with sample size of $N_e = 10^4$ elements. This sample size is practically free of sampling noise. This allows for the computation of accurate estimates of the associated pdf's, and for the background ensemble mean and covariance to be virtually identical to the real ones, i.e. $\bar{\mathbf{x}} \to \boldsymbol{\mu}^b $ and $\mathbf{P}^b \to \mathbf{B}$. %

The ensemble background covariance $\mathbf{P}^b$ is crucial since it spreads information from observed to unobserved variables, see Appendix \ref{sec:appendix} for details. %
In Section \ref{sec:real_data}, where we perform offline DA with real measurements, we estimate the background covariance from an ensemble reanalysis model product. %
In the current synthetic example, however, we prescribe a background error variance which is constant at all vertical levels, such that $\mathbf{B}$ can simply be written as a product of a common variance (scalar) and a correlation matrix:
\begin{equation}
\mathbf{P}^b = \left( \sigma^{b} \right)^2 \mathbf{C}.
\end{equation}

\noindent
We set the variance to $ \left( \sigma^{b} \right)^2  = \left( 10 \, \text{m/s} \right)^2 $. We prescribe two correlation matrices  $\mathbf{C} \in \mathcal{R}^{N_z \times N_z}$ with only positive correlations. The ${ij}^{th}$ elements of these matrices are:
\begin{equation}
\begin{aligned}
c_{i,j} &= \delta_{i,j}  \\
c_{i,j} &= \text{exp} \left( -|i-j| \right).
\end{aligned}
\end{equation}

\begin{figure}[H]
   \centering
       \includegraphics[width=0.45\textwidth]{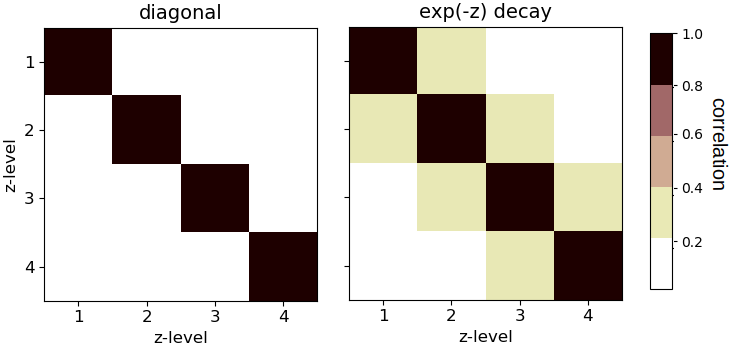}
   \caption{Two prescribed background covariance matrices for our experiments with synthetic data: a diagonal matrix (left) and a Toeplitz matrix with exponential decay in the off-diagonal elements (right). The background covariance matrix communicates observations from observed to unobserved variables. }
\label{fig:Cmatrices}
\end{figure}

\noindent
In the first case $ \delta_{i,j} $ is the Kronecker delta, so $\mathbf{C}$ becomes the identity matrix. %
The second case renders a Toeplitz matrix with a main diagonal of ones and an exponential decay for the off-diagonal elements. %
Both matrices are plotted in Figure \ref{fig:Cmatrices} for visualisation. %

We also prescribe the altitude-dependent weights $\boldsymbol{\alpha} \in \mathcal{R}^4$ applied in \eqref{eq:wc_vector}. We consider three cases:
\begin{equation}
\boldsymbol{\alpha}_{low} = \begin{bmatrix}
1 \\
0 \\
0 \\
0
\end{bmatrix}, ~~
\boldsymbol{\alpha}_{all} = \begin{bmatrix}
1/4 \\
1/4 \\
1/4 \\
1/4
\end{bmatrix}, ~~
\boldsymbol{\alpha}_{top} = \begin{bmatrix}
0 \\
0 \\
1/2 \\
1/2
\end{bmatrix}.
\label{eq:hssimple} 
\end{equation}

\noindent
For $\boldsymbol{\alpha}_{low}$, the effective cross-wind simply becomes the cross-wind at the lowermost level, %
while for $\boldsymbol{\alpha}_{all}$, the effective cross-wind becomes the averaged cross-wind over all four altitude levels. %
For $\boldsymbol{\alpha}_{top}$, the effective cross-wind is the average of the two cross-winds at the highest levels. %
This case is less realistic, but included for comparison. %

The celerity is set to the fixed value $\celerity=300$\,m/s, and we assimilate an observation with a given value and the prescribed uncertainty:
\begin{equation}
\Delta \theta  = 0.2~\text{rad}, ~\sigma^o = 0.02~\text{rad}.
\end{equation}

\noindent
Solving from \eqref{eq:obs_eq} yields $\mathbf{w}^c \approx 60 \pm ~6.2$\,m/s. We find this approximate corresponding error using the linear approximation: $\sigma^{b \leftarrow o} \approx \celerity \sigma^{o}$. 

\begin{figure}[H]
   \centering
       \includegraphics[width=0.8\textwidth]{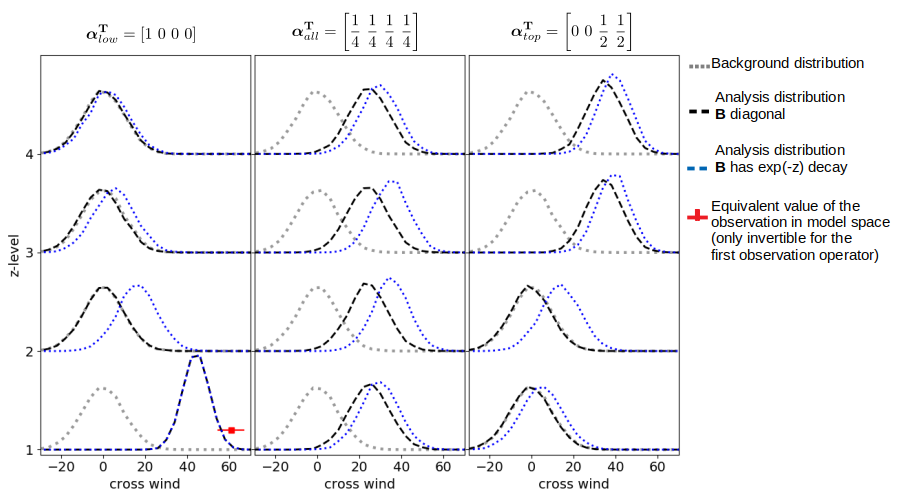}
   \caption{Synthetic-data experiments to illustrate the assimilation of one observation into four atmospheric DA vertical levels with two different background covariances (see legend) and three different vertical coefficients (panels). We use a 10000-member ensemble to avoid sampling error.}
\label{fig:example4}
\end{figure}

Figure \ref{fig:example4} shows the results of the assimilation experiments considering the two matrices $\mathbf{P}^b$ given above. %
This figure has 3 columns, one for each set of weights $\mathbf{\alpha}$. We plot several pdf's in each panel. To ease visualisation, the pdf's are scaled, and hence the vertical axes have no units. The background pdf estimated from the model ensemble is shown with a grey dotted line for the four vertical levels and operators. We also plot the analysis pdf's for the two covariances. %
When $\mathbf{P}^b$ is diagonal, the DA process can only update the levels with non-zero values in $\mathbf{\alpha}$. The analysis pdf's corresponding to this case are shown by black dashed lines. %
In the left column, only the lowest level is updated, while in the centre column the 4 levels are updated. %
In the right column, only the top two levels are updated. All observed levels are updated similarly as we apply a non-zero operator with equal values. 

A non-diagonal covariance matrix $\mathbf{P}^b$ yields a different result because non-zero off-diagonal values communicate information from observed to unobserved levels. %
The blue dotted lines show the analysis pdf's for this case. %
The magnitude of the update decreases with distance between the observed layers and non-observed layers, as expected from the exponential off-diagonal decay in $\mathbf{P}^b$.

\section{\label{sec:real_data}Offline assimilation experiments using observed infrasound from explosions}
We finally proceed to perform offline DA based on real infrasound recordings. The offline character implies that the assimilation at a given observed time is independent from all other times.

\subsection{Observations and background}
Our observations come from a dataset recording explosions at the Hukkakero site in northern Finland \citep{Gibbons2007,Gibbons2015,gibbons2019characterization,blixt2019estimating}. %
These explosions series are conducted during August and September, with individual explosions typically separated by about 24 hours. %
The dataset considered in the current study covers the years 2001--2018. %
The infrasound waves produced by these explosions are detected at the ground-based ARCES array station in Norway, which is located 178\,km due north from the explosion site. %
It takes the wave around 10 minutes (in average) to propagate from the source to the station. %

Since we know the exact explosion and detection times, as well as the exact source and receiver locations, we can compute the celerity $\celerity$ value with high accuracy. In fact, we will consider it to be error-free. The backazimuth deviation angle $\Delta\theta$ for each explosion is obtained from observations. For these observations we consider an unbiased error following a normal distribution with a standard deviation of $1/20$ of a degree. See \cite{blixt2019estimating} or, e.g., \cite{Szuberla2004} for details on the estimation of observational error in this case.

\begin{figure}[H]
   \centering
       \includegraphics[width=0.950\textwidth]{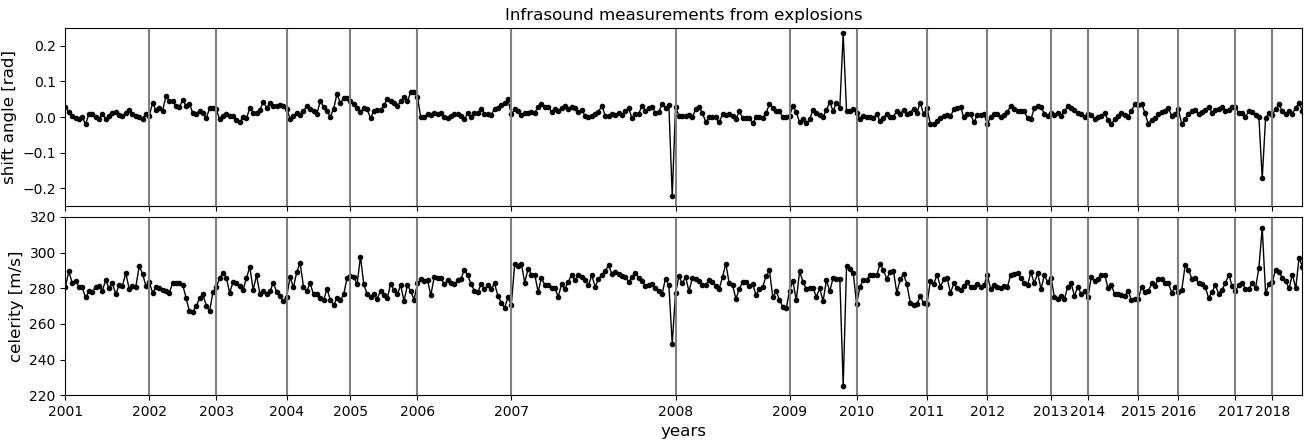}
   \caption{Time series for the observations of backazimuth deviation (top) and celerity (bottom) for the explosion events in Finland, as detected in Norway. Vertical lines separate different years and the vertical extent of each year section reflects that the number of explosions varies. The exact time of each event is not depicted for ease of visualisation.}
\label{fig:delta_cel_years}
\end{figure}

Figure \ref{fig:delta_cel_years} displays the backazimuth deviation $\Delta\theta$ (top panel) and the celerity $\celerity$ (bottom panel) for each explosion. %
The years are separated by black vertical lines. To facilitate visualization we do not display the exact time of each explosion. We discard data points where the magnitude of the backazimuth deviation is $\vert \Delta\theta \vert \ge 0.75 ~\text{rad}$ (not shown in the figure), retaining a total of $N=370$ valid events. Table \ref{tab:data} lists the number of events used and discarded for each summer.

We extract the background cross-winds from the ERA5 reanalysis product \citep{hersbach2019era5}, which has 10 ensemble members. %
We interpolate the horizontal winds from the native grid to the along-track and cross directions to the great circle connecting Hukkakero and ARCES. This is done for all the 137 ERA5 vertical levels. The time resolution of ERA5 ensemble product is 3 hours, so we linearly interpolate the wind values to the origin time of the explosion. The propagation time from source to receiver, which is around 10 minutes, is disregarded when extracting the ERA5 winds. This simplification would not be valid for longer propagation times. 

\begin{figure}[H]
   \centering
       \includegraphics[width=1.0\textwidth]{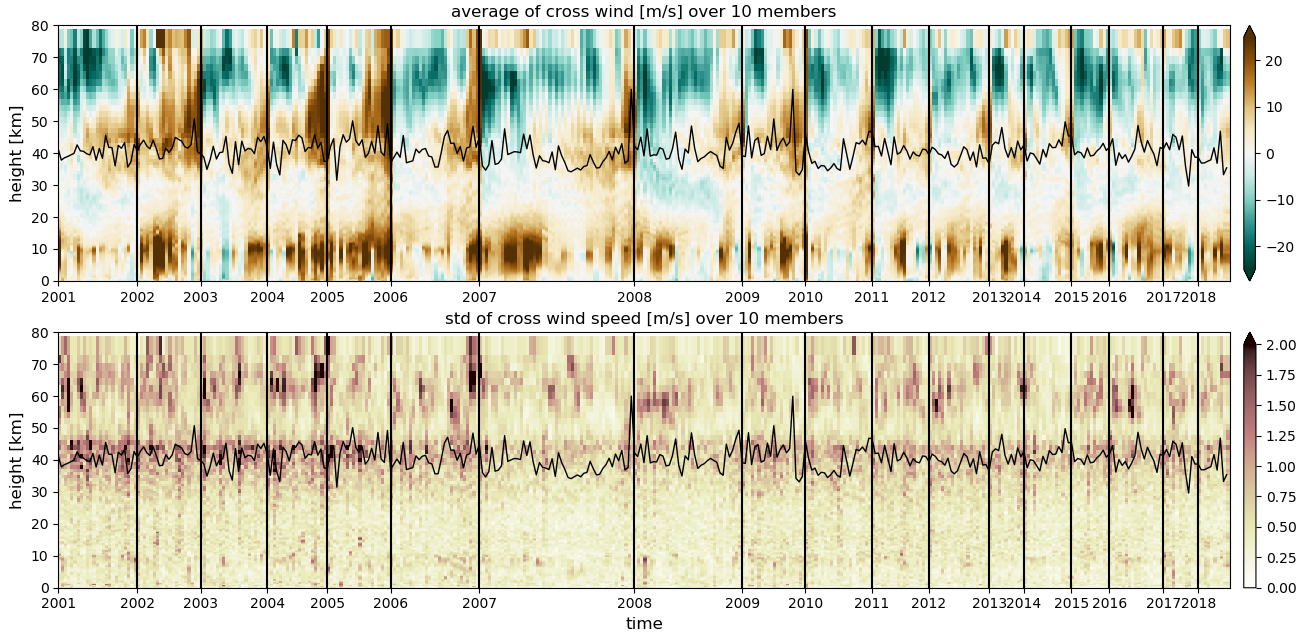}
   \caption{Cross-wind values coming from ERA5 which we use as background for our assimilation experiments (interpolated). The top panel displays the 10-member ensemble mean, and the bottom panel the 10-member standard deviation for each one of the explosion events (horizontal axis) and each one of the 137 vertical levels (which are not equally spaced). The vertical black lines separate different years, and the time-evolving line centred around 40\,km shows the maximum vertical penetration of the infrasound waves.}
\label{fig:bgd_years}
\end{figure}

Figure \ref{fig:bgd_years} shows statistics for the background cross-wind velocities for the 137 vertical levels (vertical axis) at the time of each explosion over the 18 years (horizontal axis). The vertical lines show the change of year and again the exact times are not shown in the axis. Note that the vertical levels do not have uniform resolution. %
The top panel displays the sample mean over the 10 ensemble members. We scale the colours to cover $W^c_{n_z} \in [ -25 , 25]$\,m/s. In general, the mean cross-wind speed is characterised by a strong positive jet in the lower levels (around $z=10$\,km), and a strong negative cross-wind in the upper levels (around $z=60$\,km). The cross-wind shows, however, a significant variation in time. 

The bottom panel of Figure \ref{fig:bgd_years} shows the cross-wind sample standard deviation over the 10 ensemble members. Lower levels have smaller standard deviations than higher levels. For instance, the region above $z=50$\,km has standard deviations of up to $2$\,m/s or larger, whereas the standard deviation in levels below $30$\,km are rarely higher than $0.5$\,m/s. %
This is expected since the reanalysis data contains information from atmospheric wind observations from these altitudes. %
The number of observations generally reduces with height \citep{DuruisseauHuret2017}. %
This plot suggests the observational impact of the infrasound measurements to be higher in the levels above around $z=30\,km$. However, the other factor for this impact involves the coefficients for different vertical levels, which is something we discuss in the next subsection. 

Figure \ref{fig:bgd_years} displays a time-varying black line at around $z=40$\,km. This represents the estimated maximum altitude the infrasound penetrates before being reflected towards ground. %
Any altitudes above those lines cannot be updated directly from the observations. Therefore, updates above this line are due to vertical covariances in the DA process. 
The return altitude of the infrasound is estimated by matching the travel time of a modelled infrasound ray through the model atmosphere with the observed infrasound travel time, as explained in \citet{blixt2019estimating}.

\subsection{Vertical weights}
\begin{figure}[H]
   \centering
       \includegraphics[width=0.50\textwidth]{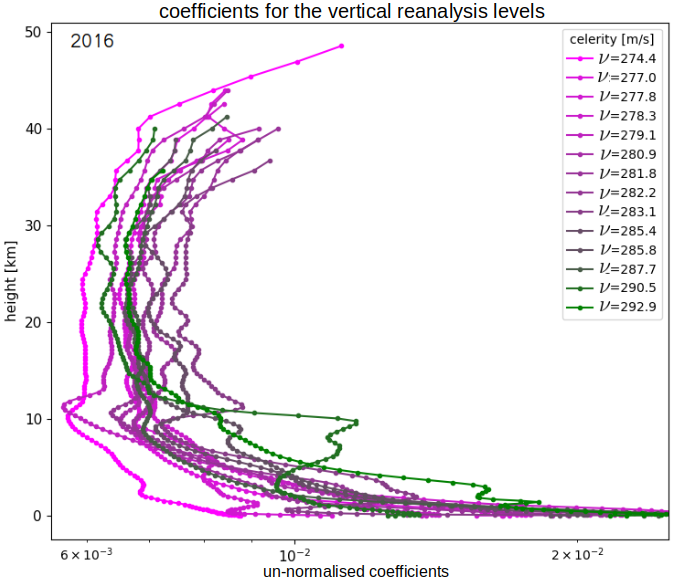}
   \caption{Non-normalised vertical weights (horizontal axis) for the 14 explosion events of 2016 as a function of the reanalysis vertical levels (vertical axis). The colours denote the values of celerity for each event (see legend).}
\label{fig:sensitivities_v2016}
\end{figure}

In the synthetic experiments we prescribed coefficients to compute the weighted cross-wind average. %
In the current section, we estimate these weights from ray-tracing through wind and temperatures \citep{blixt2019estimating} extracted from the ERA-Interim reanalysis atmospheric product \citep{dee2011_era-interim}. This is shown in Figure \ref{fig:sensitivities_v2016} for 14 events in 2016. %
The lines are coloured according to the corresponding celerity $\celerity$ as indicated in the label box on the right of the figure. %

This figure shows un-normalised vertical weights $\hat{\alpha}_{n_z}$ for each explosion. 
Notice that none of the explosion-generated infrasound waves penetrate higher than $50$\,km altitude, with the majority only reaching around $40$\,km. %
It is clear that the waves spend a significant part of the propagation time within the lowermost $10\,$km levels and within $30$ and $40$\,km. The celerity $\celerity$ ranges between $\celerity=274.4\,$m/s and $\celerity=292.9\,$m/s for these events.  

\begin{figure}[H]
   \centering
       \includegraphics[width=0.950\textwidth]{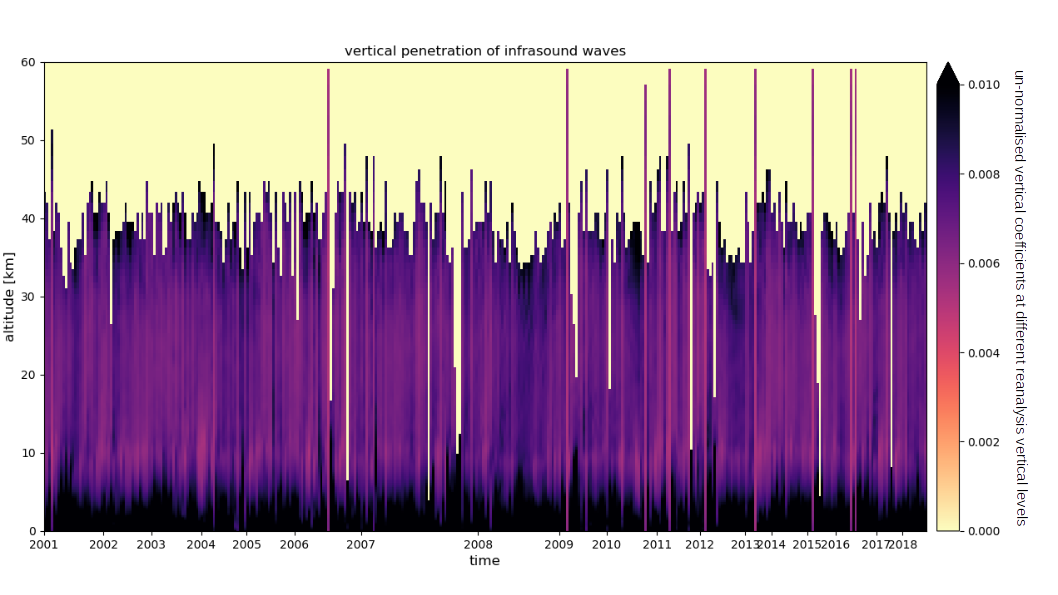}
   \caption{Vertical weights for the infrasound waves for the explosion events from 2001 to 2018 for the reanalysis vertical levels. The horizontal axis does not show the exact times for the events, only the change of year. Note that zero values correspond to altitudes where waves have not penetrated.}
\label{fig:sensitivities_vall}
\end{figure}

This process is applied to all 370 explosions, yielding vertical coefficients and maximum vertical penetration values for all the events for 18 years. %
These profiles are plotted in Figure \ref{fig:sensitivities_vall}. The horizontal axis corresponds to time, the vertical axis to altitude, and the colours correspond to the un-normalised coefficients. 
 
\subsection{The data assimilation}
To perform DA, we first need to define the vertical levels to use in the process. If we estimate the cross-wind at each reanalysis level, the size of the state variable becomes $N_x=137$. This problem is quite challenging, especially since we only have $N_y=1$ observation containing integrated information. An extra complication comes from our relatively small ensemble size $( N_e=10 )$. %
A way to simplify the problem is to create fewer DA vertical levels by applying vertical averaging. After trying several averaging kernel heights, we decided to use $N_{z\text{DA}}=6$ equidistant DA levels with a height of $\Delta Z_\text{DA} = 12$\,km, covering the altitudes $z=0$\,km to $z=72$\,km. In a given DA level $l$ with $N_{zl}$ non-equidistant reanalysis levels inside it, the vertically averaged cross-wind is:
\begin{equation}
W^c_l = \frac{ \sum_{n_z=1}^{N_{zl}} W^c_{n_z} \Delta z_{n_z} }{\Delta Z_\text{DA}}.
\end{equation}

\noindent
We use this weighted approach also to obtain a weight $\alpha_{l}$ for each DA level. We also ensure the sum of the weights to be normalised one. Starting from the un-normalised weights $\hat{\alpha}$ at native levels coming from the ray tracing, we compute: 
\begin{equation}
\alpha_l = \frac{ \sum_{n_z=1}^{N_l} \hat{\alpha}_{n_z} \Delta z_n }{\Delta Z_\text{DA} },~~\sum_{n_z=1}^{N_z}{\alpha_{n_z}}=1 \label{eq:weights_6}
\end{equation}

\begin{figure}[H]
   \centering
       \includegraphics[width=0.95\textwidth]{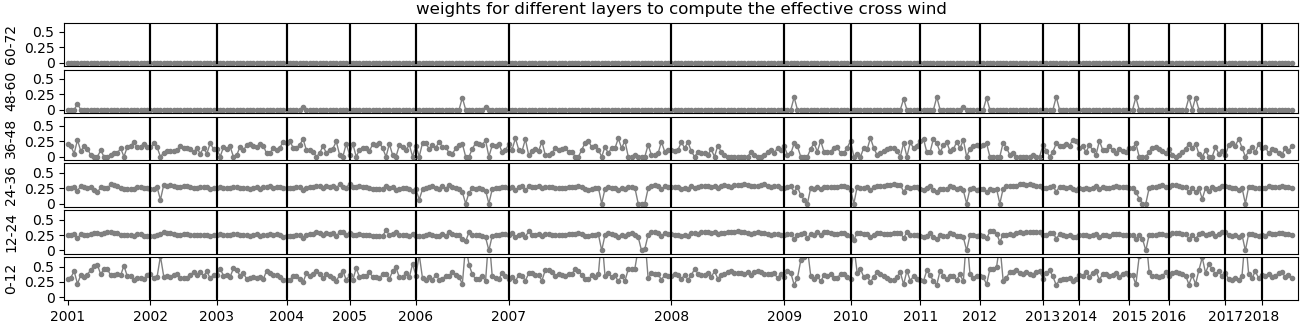}
   \caption{Time evolution of the normalised weights for the different DA vertical levels. These waves are used to compute the effective cross-wind at any time. Most waves only reach the four bottom DA levels. The top level is never reached by the infrasound wave.}
\label{fig:vert_weis}
\end{figure}

The normalised weights computed using \eqref{eq:weights_6} are plotted in Figure \ref{fig:vert_weis} for all the events (horizontal axis) and each DA vertical level. %
There is temporal variability in the weights, especially for the lowermost four levels. %
Note that the uppermost level (60--72\,km) always has zero weights, and in the next level (48--60\,km) infrasound waves penetrate for only few events per year. These upper levels can only be affected by observations through the sample covariance between different levels.

\subsection{The quality of the background covariance}
\begin{figure}[H]
   \centering
       \includegraphics[width=0.750\textwidth]{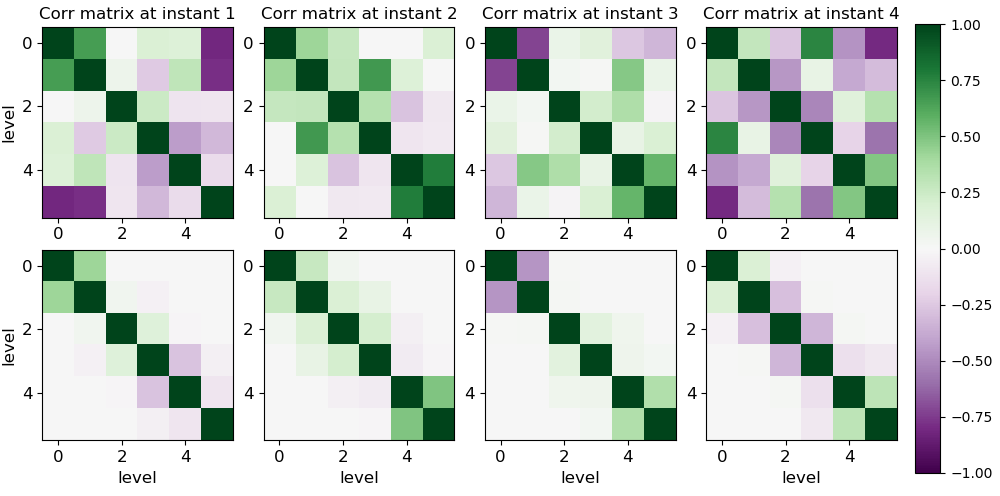}
   \caption{Sample background correlations for four different instants and the six DA vertical levels, as computed from the 10-member ERA5 reanalysis dataset. We depict both raw correlations (top row) and localised correlations (bottom row).}
\label{fig:corr_noloc_loc}
\end{figure}

An accurate representation of the background covariance matrix is vital to the DA process. Recalling that we have a limited-size ensemble, $N_e=10$, a low-quality estimator can be harmful to the analysis values. Localisation can handle poor-quality long-distance covariances when working with small ensemble sizes. %
In fact, even after reducing the problem to $N_{z\text{DA}} = 6$ DA vertical levels, we are still left with noisy background error covariance matrices $\mathbf{P}^b \in \mathcal{R}^6$. %
This is illustrated in the top row of Figure \ref{fig:corr_noloc_loc}, which shows the raw correlation matrices from the ensemble at 4 different times (columns). Green colours are positive correlations and purple colours are negative correlations. These matrices are different because they contain flow-dependent information. %
We display correlations and not covariances because the variances for different altitudes have different orders of magnitudes. %
Intuitively, the correlations should decrease with increasing vertical distance. %
The second row shows the covariances after being localised, which means they have been multiplied (using Schur product denoted $\circ$) times a matrix $\boldsymbol{\Phi}$ of coefficients that decay with distance \citep{Hamill_et_al2001}:
\begin{equation}
\mathbf{P}^b_{loc} = \boldsymbol{\Phi} \circ \mathbf{P}^b.
\end{equation}

\noindent
We apply a Gaspari-Cohn localisation function -- which is a compact-support approximation to a Gaussian \citep{Gaspari_Cohn_1999} -- with a half-width of 15\,km. We choose this to be larger than the height of each DA vertical level (12\.km).  

Another technique to improve sample covariances is inflation \citep{Anderson_and_Anderson_1999}. In its simplest implementation the ensemble of background perturbations is scaled:
\begin{equation}
\begin{aligned}
\hat{\mathbf{X}}^b_{inflated} &=  \left( 1+\rho \right) \hat{\mathbf{X}}^b \\
\therefore ~ \mathbf{P}^b_{inflated} &= \left( 1+\rho \right)^2 \mathbf{P}^b.
\end{aligned}
\end{equation}

\noindent
Inflating the background covariance increases the Kalman gain, which makes the observations have a larger impact in the analysis field. %
This makes intuitive sense: if the uncertainty in the background is considerably larger than the uncertainty in the observations, the assimilation should tend to ignore the background. %
There exist more advanced inflation implementations, e.g. where time and space-dependent coefficients are applied  \citep{Miyoshi_2011, Raanes_et_al_2019}, but in the current experiment these are fixed. %
Moreover, it is common to choose a $\rho$ value which minimises a set of accuracy measures -- e.g. the root mean squared error of the analysis mean -- with respect to independent observations. In the current work, we do not have reliable estimates of such verification values, so instead we study the impact of different inflation values.

A clear reason for applying inflation is that the ensemble background covariance is often underestimated. This is inherent to small ensemble sizes,  see \cite{vanLeeuwen_1999, Sacher_Bartello_2008, Amezcua_vanLeeuwen_2018} for detailed explanations of direct and indirect effects. There are more tangible mechanisms for the misrepresentation of the background covariance. This includes differences in the resolution of model and observations, and the imperfect representation in the forecast and observational process. %

In our experiment setup we recognise there are sources of imperfection. These include (a) we temporally interpolate from the reanalysis times to the time of the observation, (b) we consider instantaneous velocities while the infrasound wave propagates for around 10 minutes, (c) there may be erroneous assumptions behind the calculation of the $\alpha$ weights inside the ray-tracing technique. %
We performed the experiments with several inflation values $\alpha$, and below we discuss the results obtained using two of these values.

\subsection{Results}
Here, we display the results for the following DA settings: $N_x = N_{z\text{DA}} = 6$ state variables per observational instant, $N_y=1$ observations, $N_e=10$ ensemble members, vertical localisation with a half-width of $15$\,km, vertical weights coming from the ray-tracing assumed to be perfect, and two different inflation factors $\rho=0,~\rho=1$. The second inflation value means the standard deviation of the background is doubled compared to the data in the non-inflated assimilation.

\begin{figure}[H]
   \centering
       \includegraphics[width=0.90\textwidth]{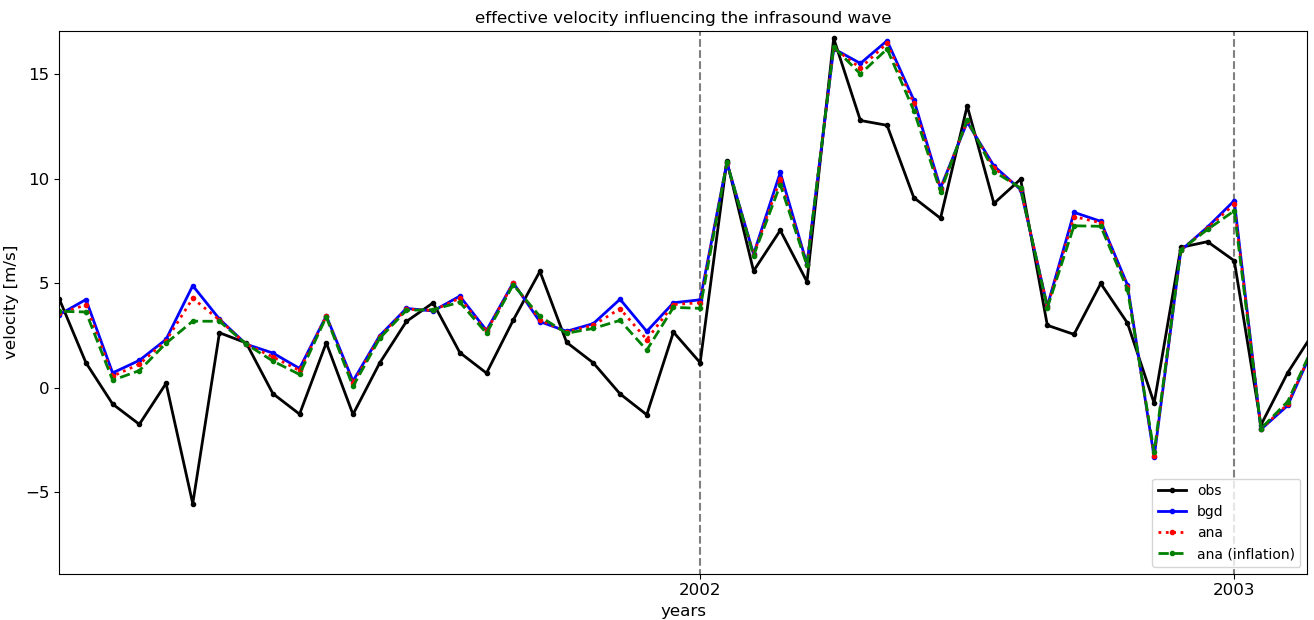}
   \caption{Effective cross-wind computed from infrasound backazimuth observations, and from the background and analysis mean values for the six DA levels. For clarity we only display a short time interval (2001-2003).}
\label{fig:weff}
\end{figure}

Figure \ref{fig:weff} shows the weighted cross-wind solved from \eqref{eq:DTheta} for observations (black line) and computed from \eqref{eq:Wc_vec} for the background (blue line), as well as the resulting analysis (red and green lines, depending on the inflation). To facilitate visualisation, we only display the years 2001 and 2002.  

The background and observation cross-winds are similar for some of the events, but for most events the DA produces changes. In fact, for some events the difference is up to $1$ or $2\,$m/s.
In the absence of inflation, the background and analysis values are quite close. The use of inflation, however, increases the differences between analysis and background as expected. 

The impact of the observations is in general low, especially in the absence of inflation. Several factors can explain this: First, the variance of the background ensemble is small, which is expected since this is a reanalysis product already containing information. The observation impact might be greater if instead using an ensemble forecast as background. Second, as already mentioned, the ensemble size is small with only $N_e=10$ members. A larger ensemble would allow to select different state variables, for instance a larger number of DA vertical levels. Less vertical averaging of the original variables would give a prior with larger variance, hence allowing for larger observational impact. It is important to point out that in an online setting, the background would come from an ensemble forecast and the infrasound observations would not be the only data assimilated. Another aspect is that the stratospheric winds are in general significantly weaker and less variable in August and September than in winter. It will be interesting, in a future work, to perform these experiments for wintertime explosions and to assess the observational impact. 

\begin{figure}[H]
   \centering
       \includegraphics[width=0.95\textwidth]{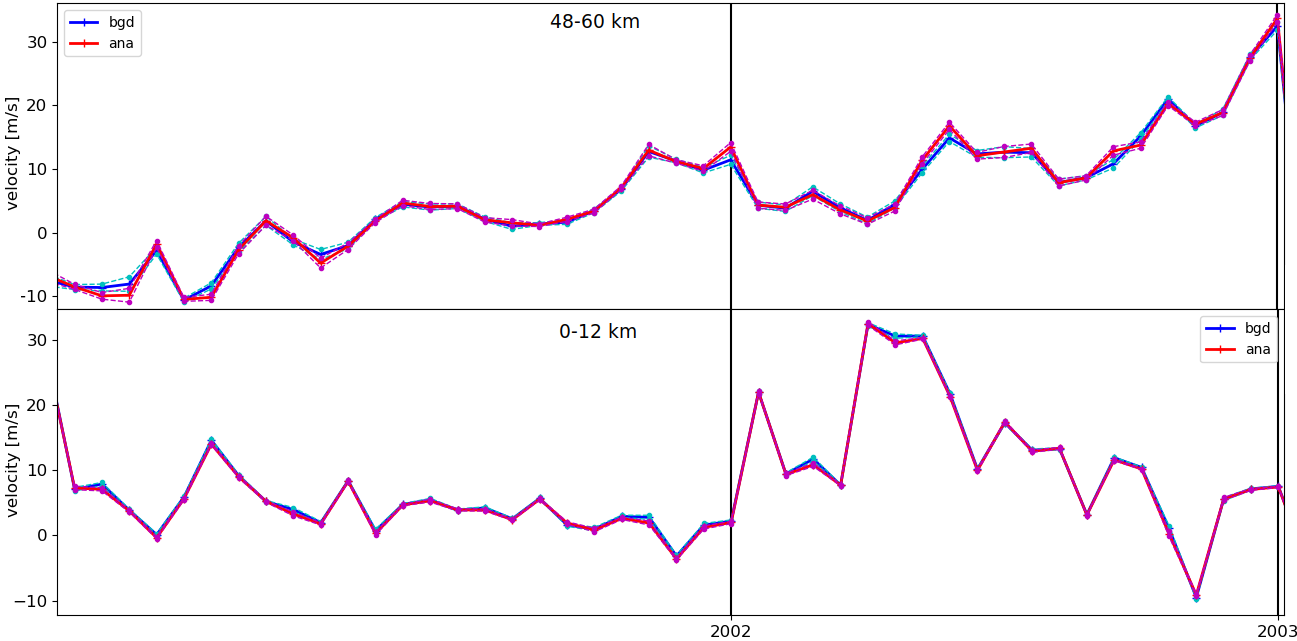}
   \caption{Background (thick blue) and analysis (thick red) mean cross-winds in two DA vertical levels: 0--12\,km (bottom) and 48--60\,km (top). Only a short time interval (2001-2003) is displayed. The analysis was obtained using inflation. The thin cyan lines show ten individual backgrounds (one for each ensemble member), and the thin magenta lines show the respective ten analysis values. }
\label{fig:assim5layers}
\end{figure}

How do changes in the vertically averaged cross-wind translate to the different DA vertical levels? These results are shown in Figure \ref{fig:assim5layers}, which has two panels corresponding to two selected vertical levels: $0-12$\,km (bottom) and $48-60$\,km (top) for the 2001 and 2002 events. %
The blue line shows the background mean, with the cyan lines to each side indicating one standard deviation. %
The red line denotes the analysis mean, with the magenta lines to each side indicating one standard deviation. This analysis was produced using inflation. %
There are some changes in the values of the cross-wind in the lower level, however these tend to be small. %
The difference between background and analysis is more noticeable at higher DA levels, which are not even updated directly (recall most explosions do not penetrate these altitudes) but based on the inter-level covariances. In the no-inflation case, there are still changes, but they are less distinguishable in the plot.

To evaluate the results for all DA events without inflation, we compute the following derived quantities at each time $t$:
\begin{subequations}
 \begin{align}
 d_{n_z}^{ab,t} &= \bar{x}_{n_z}^{a,t} - \bar{x}_{n_z}^{b,t} \label{eq:innov}, \\  
 r_{n_z}^{ab,t} &= \left( \frac{ s_{n_z}^{a,t} } { s_{n_z}^{b,t} } \right)^2. \label{eq:ratio} 
 \end{align}
\end{subequations}

\begin{figure}[H]
   \centering
       \includegraphics[width=1.00\textwidth]{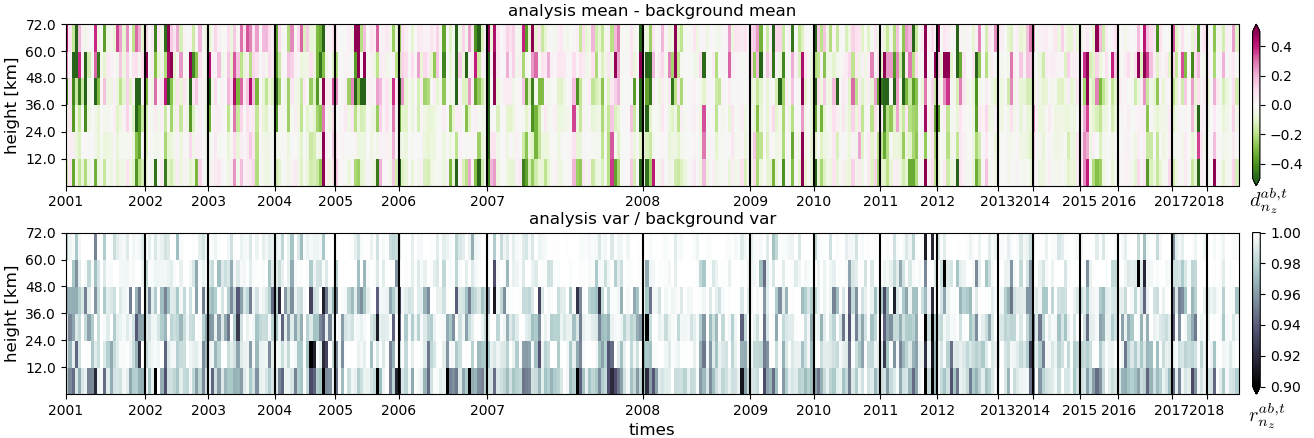}
   \caption{Results of the DA process as a function of time (horizontal axis, with vertical black lines denoting the change of year) and DA vertical level (vertical axis). The top panel displays the analysis increment, i.e. the difference of the analysis mean minus the background mean. The bottom panel shows the ratio of analysis variance divided by background variance. No inflation was used in the experiments illustrated in this figure.}
\label{fig:innovs_ratios}
\end{figure}

\noindent
The quantity \eqref{eq:innov} is called \emph{analysis increment}, which is the difference between the analysis mean and the background mean. %
The second quantity \eqref{eq:ratio} is a variance ratio, which is the analysis variance divided by the background variance. Both the analysis increment and the variance ratio are computed for each DA level and for each observation time. %
These quantities are displayed in Figure \ref{fig:innovs_ratios} for all events as a function of time (horizontal axis) and altitude (vertical axis). The DA process does not only result in a modified mean, but it also reduces the uncertainty of the estimate. %
In mathematical terms, the trace of the analysis covariance matrix is smaller than the norm of the background covariance (see e.g. \cite{Asch_et_al2016}). %

The top panel displays the innovations with typical magnitudes between $-0.2$\,m/s and $0.2$\,m/s. Pink colours represent positive increments and green colours represent negative increments. Remember that these increments are the changes that the infrasound observations produce to the forecast. The bottom panel displays the resulting variance ratios of standard deviations. The figure plot confirms that these values, as expected, are always smaller than $1$. %
The darkest colours correspond to the greatest reduction in uncertainty. The experiment results in a ratio which descends to around $0.9$. However, we keep in mind that the reanalysis data ensembles already contained small statistical uncertainty.  

\begin{figure}[H]
   \centering
       \includegraphics[width=0.5\textwidth]{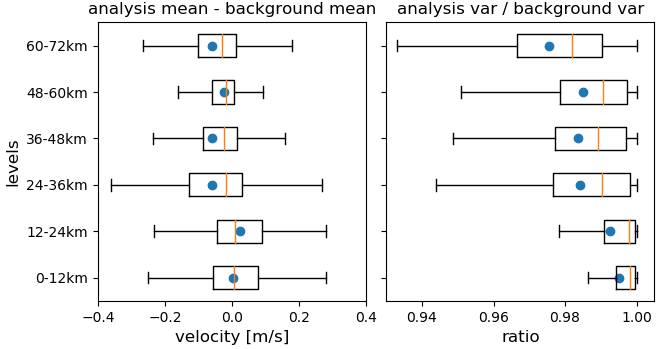}
   \caption{Summary of the mean analysis increment $d^{ab}$ (left) and variance ratio $r^{ab}$ (right) for each vertical level. The values are summarised using boxplots, where the outliers are omitted for clarity in the figure. The mean values are shown with the blue dots, while the median is shown with orange vertical lines. This figure shows that the observations are having an impact, which albeit small, is non-zero. The largest observation impact is in the $24-36$\,km DA vertical level.}
\label{fig:innovsbp_years}
\end{figure}

Figure \ref{fig:innovsbp_years} summarises the increment $d^{ab}$ (left) and the variance ratios $r^{ab}$ (right) obtained, for each vertical level. Box plots provide a non-parametric summary (with outliers omitted to avoid cluttering the figure). These box plots are complemented by the mean as shown by blue dots. %
There are non-zero innovation results at all vertical levels, with the largest typically within the level $24-36$\,km, and the smallest typically within $48-60$\,km altitude. %
In the three upper-most altitude layers, at least $75\%$ of the increments are negative. 
Note that in at least three levels, the mean and the median differ significantly, indicating asymmetry in the distribution of the innovations.

The right panel of Figure \ref{fig:innovsbp_years} shows the resulting variance ratio, which is 0 and 1. Since this has a non-symmetric distribution, the mean and median does not coincide. %
Notice that the reduction of the variance is largest in the four upper levels, i.e.\ $24-72$\,km. %
This is expected because these levels have greatest background uncertainty. Since the waves penetrate only to around $40$\,km, updating the DA levels at these altitudes is done both directly and via covariances. %
The lowermost two levels exhibit a limited covariance reduction. %
Although the coefficients for the lowest DA level are significant, the background winds are already well constrained there, hence only allowing the assimilated infrasound-based data to impact the analysis to a minor extent.

\section{\label{sec:conclusions}Summary and future work}
This is the first study to explore assimilation of atmospheric infrasound data into atmospheric models in order to constrain atmospheric winds. %
The backazimuth deviation of infrasound waves carries integrated information related to the cross-winds acting on the wave along its atmospheric propagation path. %
We show that assimilation of this information using an ensemble Kalman Filter is able to provide corrections to the wind in stratospheric and tropospheric altitudes. 

We performed DA experiments for 370 explosion events throughout 18 years (2001--2018). %
We know the accurate time and location of the explosions and arrivals of the infrasound waves. This allows us to accurately calculate the propagation time and the celerity $\celerity$. It also permits performing complementary ray-tracing to determine the vertical sensitivity at different vertical levels, which is needed in the observation operator. %
To reduce the dimensionality of the problem, we consider average values corresponding to $N_z=6$ DA levels, each 12\,km thick. This is opposed to the original $N_z=137$ levels of the reanalysis. Here, there might be room for improvements and subsequent works can explore in detail the effect of selecting different numbers of DA vertical levels. 

The results of the DA experiments yielded non-zero analysis increments (defined as analysis mean minus background mean) for most times, with the largest values in the $24-36$\,km layer. %
More than $75\%$ of the increments calculated above $36$\,km are negative, suggesting a bias in the background values. %
As required by construction, the variance in the cross-wind values at all levels has been reduced for all data points assimilated, while for the upper-most levels the reduction reaches up to $2-5\%$. This implies a reduction of the uncertainty in the estimation.

It would be desirable to apply this framework to existing datasets for explosions performed during the winter season when the stratosphere is more dynamic than in August and September. This may present a challenge, though, since larger magnitudes of cross-wind can reduce the linearity of arc-tangent in  \eqref{eq:DTheta}. This may present a challenge to the DEnKF. We can instead try with techniques that handle departures from linearity better. For instance, the iterative ensemble Kalman smoother (e.g. \citep{Evensen_etal2019} is a useful candidate to solve this problem.

For future work, we suggest exploiting signals from natural continuous sources like microbaroms. These are atmospheric infrasound waves produced by ocean surface hot-spots where counter-propagating surface waves are prevalent \citep{posmentier1967theory,donn1971natural,LePichon_etal2006,denouden2020,decarlo2020}. 

In this work we rely on many simplifications. In the future we aim to solve a setup akin to the middle panel of Figure \ref{fig:geometry}. Namely, this would consider the cross-wind variation both on the vertical and the along-track direction of the infrasound wave. This becomes especially important when the distance between source and receiver increases. An example is the detection in Norway of infrasound from microbaroms generated near Iceland; in this case the separation is about 2000\,km and considering a single horizontal slab may be detrimental to the usefulness of the estimation. In this case we may also not be able to consider the winds constant in time for each position along the trajectory.

We have an important advantage when working with the explosions data set: the times and locations of both the emission and detection of the infrasound waves are known accurately. This, in turn, allows us to consider the celerity $\celerity$ perfectly known, which we have done in this work. In the case of microbaroms, for instance, the time and location of the detection may be accurately known, but the location and time of the emission may prove much more elusive. In these cases, there may be large uncertainty on the values of celerity. This, added to the uncertainty in the propagation medium, makes it necessary to consider celerity as another random function. Several previous works establish the pdf for celerity in infrasound propagating under stratospheric waveguide conditions. For example, \cite{blom2015improved} used simulations to establish the expected celerity to be between 250 and 350\,m/s for propagation distances at around 200\,km. Similarly, \cite{morton2014celerity} analysed both simulations and measurements to find a celerity distribution at 275\,km distance with values between around 280 and 310\,m/s. A data-based study presented in \cite{Nippress2014} estimates the celerity distribution at 200\,km distance to span the 270 to 300\,m/s range.  

Regarding the dynamics of the wave propagation, we recognise that the framework applied here requires an auxiliary ray-tracing method to determine the sensitivity to the wind at different vertical levels (first native and then DA levels) in the weighted sum giving the average cross-wind impacting the observation. Follow-up studies could include the development of approaches to instead estimate these sensitivity weights as part of the assimilation process. Then the implementation of an expression akin to \eqref{eq:alpha_spatial} might be required. In turn, this would require the state variable to include the along-track wind and the temperature (which the sound speed is a function of) in each of the grid points traversed by the wave. However, this would also give an opportunity to estimate along-track winds. 

An important detail to mention is that a DA process requires a verification step to assess the quality of the analysis field obtained. %
For identical-twin experiments, one produces a synthetic truth from which the simulated observations were extracted. Then the analysis can be assessed with respect to this reference truth.  %
In operational DA the true state of the system is unknown, so verification becomes more elusive. One option is to have independent observations or independent reanalysis data that can verify the analysis. %
In the current study, we do not have independent observations for validation. This in turn restricted us from tuning the values of localisation radius and inflation parameter. %
Although this is outside the scope of the current study, prospective future studies might have access to independent measurements to allow for tuning and verification. %
Here, the ADM-Aeolus satellite mission will likely be a reliable benchmark for winds up to 30\,km altitude \citep{tan2008adm}. Likewise, future validation may be possible using data from portable lidars; for example, the CORAL system \citep{kaifler2017observational,kaiflers2015} might be upgraded to provide direct wind measurements. %

Finally, the DA experiments of this study were made offline. In order to perform online assimilation experiments, it would be necessary to implement the methodology in a dynamic forecasting system. In an operational or quasi-operational setting, infrasound measurements can be added to the rest of the available observations at the moment of assimilation. Although implementation in an operational assimilation system still requires substantial further work, the methodology described in the present study provides a starting point for such developments. This an objective of a next step following up the European ARISE and ARISE2 projects \citep{blanc2018toward, blanc2019middle}. 

Given that single-station infrasound measurements provide atmospheric wind measurements within a sparsely observed altitude range for a given geographical region, an extended or even global multi-station wind sampling might be feasible using, e.g. infrasound station data recorded by the International Monitoring System network \citep{dahlman2009nuclear, marty2019ims}. %
Hence, there are several opportunities yet to explore in further work related to atmospheric probing and data assimilation using infrasound datasets. 
A long-term objective is to enhance or constrain the representation of stratospheric winds in global models, thereby contributing to enhanced surface weather predictions on subseasonal-to-seasonal timescales \citep{Domeisen_etal_2019a,Domeisen_etal_2019b}.

\section*{Acknowledgments}
We thank two anonymous reviewers whose insightful comments and suggestions helped improve and clarify this manuscript. JA acknowledges support and funding from the UK National Centre for Earth Observation. %
This work was supported by the project \emph{Middle Atmosphere Dynamics: Exploiting Infrasound Using a Multidisciplinary Approach at High Latitudes} (MADEIRA), funded by the Research Council of Norway basic research programme FRIPRO/FRINATEK under contract No.\,274377. SPN and EMB also acknowledge NORSAR institute funding. %
This study was facilitated by previous research performed within the framework of the ARISE and ARISE2 projects  \citep{blanc2018toward, blanc2019middle} funded by the European Commission FP7 and Horizon 2020 programmes (grant agreements 284387 and 653980). %
The authors declare no conflicts of interest.

\newpage
\section{Figures}

\section{Tables}
\begin{table}[H]
\centering
 \rowcolors{4}{}{gray!10}
 \begin{tabular}{@{}p{0.06\columnwidth}<{\centering}@{}p{0.1\columnwidth}<{\centering}@{}p{0.1\columnwidth}<{\centering\arraybackslash}@{}}
 &\multicolumn{2}{c}{Number of observations}\\
 \cmidrule(lr){2-3}
 Year & Included &Discarded\\
 \toprule
 2001 &  26  &  0\\ 
 2002 & 20 & 0 \\
 2003 & 21 & 0 \\
 2004 & 19 & 1 \\
 2005 & 20 & 1 \\
 2006 & 28 & 0 \\
 2007 & 49 & 0 \\
 2008 & 34 & 1 \\
 2009 & 20 & 1 \\
 2010 & 21 & 1 \\
 2011 & 18 & 1 \\
 2012 & 19 & 2 \\
 2013 & 11 & 0 \\
 2014 & 15 & 0 \\
 2015 & 12 & 0 \\
 2016 & 17 & 2 \\
 2017 & 11 & 0 \\
 2018 & 9 & 0 \\
 \midrule
 Total & 370 & 10 \\ 
 \bottomrule
\end{tabular}
\caption{Yearly number of included and discarded infrasound explosion observations in the assimilation experiments. } 
\label{tab:data}
\end{table}

\bibliographystyle{abbrvnat}
\bibliography{infrasound_lib.bib}

\appendix
\section{The ensemble Kalman Filter framework applied in this study\label{sec:appendix}}
In this work we use the Deterministic Ensemble Kalman Filter \citep{sakov_oke_denkf2008}. The Kalman filter \citep{kalman1960,Kalman61newresults} is a minimum-variance DA algorithm which relies on the mean and covariance of the state variable. It has two steps: forecast and analysis. It is optimal under Gaussian statistics for the sources of additive error, and linear observation and evolution operators. Under these conditions the process yields a full Bayesian solution of the problem (see, e.g.\ \cite{Asch_et_al2016}) .

We perform offline experiments (no forecasts), hence we focus the explanation on the analysis step. Let the state variable (at any time) $\mathbf{x} \in \mathcal{R}^{N_x}$ follow a Gaussian distribution with $E \left( \mathbf{x} \right) = \boldsymbol{\mu}^b \in \mathbf{R}^{N_x}$ and $Cov \left( \mathbf{x} \right) = \mathbf{B} \in \mathbf{R}^{N_x \times N_x}$. 
An observation $\mathbf{y} \in \mathcal{R}^{N_y}$ is obtained as:
\begin{equation}
\mathbf{y} = \mathbf{H} \mathbf{x}_{true} + \boldsymbol{\eta}.
\label{eq:lin_obs}
\end{equation}

\noindent
$\mathbf{H} \in \mathcal{R}^{N_y \times N_x}$ is the observation matrix and $\boldsymbol{\eta} \in \mathcal{R}^N_y$ is the observation error with expected value $E \left[ \boldsymbol{\eta} \right] = \mathbf{0}$ and covariance $Cov \left[ \boldsymbol{\eta} \right] = \mathbf{R} \in \mathcal{R}^ {N_y \times N_y} $. Usually $\mathbf{H}$ loses information since often $N_y \ll N_x$.

Information from background and observations is combined via the Kalman analysis equations for mean and covariance:
\begin{subequations}
\begin{align}
\boldsymbol{\mu}^a &= \left( \mathbf{I} - \mathbf{K} \mathbf{H} \right) \boldsymbol{\mu}^b + \mathbf{K} \mathbf{y}  \\
\mathbf{A} &= \left( \mathbf{I} - \mathbf{K} \mathbf{H} \right) \mathbf{B},
\end{align}
\end{subequations}

\noindent
where $b$ and $a$ stand for background and analysis respectively. The Kalman gain $\mathbf{K}$ is:
\begin{equation} 
\mathbf{K} = \mathbf{B} \mathbf{H}^\mathbf{T} \boldsymbol{\Gamma}^{-1}.
\end{equation}

\noindent
where $\boldsymbol{\Gamma} \in \mathcal{R}^{N_y \times N_y}$ is the total covariance in observation space:
\begin{equation} 
\boldsymbol{\Gamma} = \mathbf{H} \mathbf{B} \mathbf{H}^\mathbf{T} + \mathbf{R}.
\end{equation}

Many systems of interest have non-linear evolution and observation operators, i.e.\ $ m : \mathcal{R}^{N_x} \to \mathcal{R}^{N_x} $ and $ h : \mathcal{R}^{N_x} \to \mathcal{R}^{N_y} $. This yields:
\begin{subequations}
\begin{align}
\mathbf{x}^{t} &= m \left( \mathbf{x}^{t-1} \right) \label{eq:nlin_evol} \\
\mathbf{y}^t &= h \left( \mathbf{x}_{true}^{t} \right) + \boldsymbol{\eta}^t. \label{eq:nlin_obs}
\end{align}
\end{subequations}

\noindent
The KF can still be implemented after linearising these operators. This is known as Extended Kalman Filter (EKF: see, e.g.\  \cite{Jazwinski1970}) and its use is complicated since it involves large Jacobian matrices.

An alternative is to use sample estimators for mean and covariance and to work with ensembles \citep{Evensen1994}. \citet{Hunt_et_al2007} nicely describe handling non-linear operators for this approach: 
Start with an ensemble of $N_e$ initial states, i.e.\ the matrix $\mathbf{X}^0 \in \mathcal{R}^{N_x \times N_e}$: 
\begin{equation}
\mathbf{X}^0 = \left[ \mathbf{x}_1^0, \cdots, \mathbf{x}_{N_e}^0 \right].
\end{equation}

\noindent
The background ensemble at time $t$ is found by applying the model to each member:  
\begin{equation}
\mathbf{X}^{t,b} = \left[ \mathbf{x}_1^{t,b} = m\left(\mathbf{x}_1^{0} \right), \cdots, \mathbf{x}_{N_e}^{t,b} = m\left(\mathbf{x}_{N_e}^0 \right) \right].
\end{equation}

\noindent
We now drop the time index. The sample background mean $\bar{\mathbf{x}}^b \in \mathcal{R}^{N_x}$ is:
\begin{equation}
\bar{\mathbf{x}}^b = \frac{1}{N_e} \sum_{n_e = 1}^{N_e} \mathbf{x}^b_{n_e}.
\end{equation}

\noindent
A matrix of background perturbations $\hat{\mathbf{X}}^b \in {N_x \times N_e}$ is computed as
\begin{equation}
\hat{\mathbf{X}}^b = \left[ \mathbf{x}^b_1 - \bar{\mathbf{x}}^b,  \cdots, \mathbf{x}^b_{N_e} - \bar{\mathbf{x}}^b \right].
\end{equation}

\noindent
which relates to the (low-rank) sample covariance matrix $\tilde{\mathbf{P}}^b$ as:
\begin{equation}
\mathbf{P}^b = \frac{1}{N_e-1} \hat{\mathbf{X}}^b  \hat{\mathbf{X}}^{b \mathbf{T}}.
\end{equation}

\noindent
The background ensemble in observation space $\mathbf{Y}^b \in \mathcal{R}^{N_y \times N_x}$ is obtained by applying the observation operator to each ensemble member of the background: 
\begin{equation}
\mathbf{Y}^b = \left[ \mathbf{y}^b_1 = h\left(\mathbf{x}^b_1 \right), \cdots, \mathbf{y}^b_{N_e} = h \left( \mathbf{x}^b_{N_e} \right) \right].
\end{equation}

\noindent
The sample mean $\bar{\mathbf{y}}^b \in \mathcal{R}^{N_y}$ is computed as:
\begin{equation}
\bar{\mathbf{y}}^b = \frac{1}{N_e} \sum_{n_e = 1}^{N_e} \mathbf{y}^b_{n_e}
\end{equation}
\noindent
and the matrix of perturbations in observation space $\hat{\mathbf{Y}}^b \in {N_y \times N_e}$ is
\begin{equation}
\hat{\mathbf{Y}}^b = \left[ \mathbf{y}^b_1 - \bar{\mathbf{y}}^b,  \cdots, \mathbf{y}^b_{N_e} - \bar{\mathbf{y}}^b \right].
\end{equation}

\noindent
This allows to compute the following expressions:
\begin{subequations}
\begin{align}
\mathbf{P}^b \mathbf{H}^{\mathbf{T}} &= \frac{1}{N_e-1} \hat{\mathbf{X}}^b \hat{\mathbf{Y}}^{b\mathbf{T}} \label{eq:PHT} \\
\mathbf{H} \mathbf{P}^b \mathbf{H}^{\mathbf{T}} &= \frac{1}{N_e-1} \hat{\mathbf{Y}}^b \hat{\mathbf{Y}}^{b\mathbf{T}} \label{eq:HPHT} \\
\tilde{\boldsymbol{\Gamma}} &= \frac{1}{N_e-1} \hat{\mathbf{Y}}^b \hat{\mathbf{Y}}^{b\mathbf{T}} + \mathbf{R}
\end{align}
\end{subequations}

\noindent
and the computation of the ensemble-based gain is simply
\begin{equation}
\tilde{\mathbf{K}} = \frac{1}{N_e-1} \hat{\mathbf{X}}^b \hat{\mathbf{Y}}^{b\mathbf{T}} \tilde{\boldsymbol{\Gamma}}^{-1}. \label{eq:K_SO}
\end{equation}

The analysis equation for the mean is simply:
\begin{equation}
\mathbf{x}^a = \mathbf{x}^b + \tilde{\mathbf{K}} \left( \mathbf{y} - \bar{\mathbf{y}}^b \right).
\end{equation}

\noindent
and the equation for the perturbations is: 
\begin{equation}
\hat{\mathbf{X}}^a = \hat{\mathbf{X}}^b \left( \mathbf{I} - \frac{ \mathbf{Y}^{b \mathbf{T}} \tilde{\boldsymbol{\Gamma}}^{-1} \mathbf{Y}^b }{N_e-1} \right)^{1/2}.
\end{equation}

\noindent
It can be computationally expensive to evaluate this matrix square root. %
\cite{sakov_oke_denkf2008} consider a Taylor expansion approximation to the first two terms, which yields a more efficient expression:
\begin{equation}
 \left( \mathbf{I} - \frac{ \mathbf{Y}^{b \mathbf{T}} \tilde{\boldsymbol{\Gamma}}^{-1} \mathbf{Y}^b }{N_e-1} \right)^{1/2} \approx \mathbf{I} - \frac{1}{2} \frac{ \mathbf{Y}^{b \mathbf{T}} \tilde{\boldsymbol{\Gamma}}^{-1} \mathbf{Y}^b }{N_e-1}.
\end{equation}

\noindent
Then, the perturbation update equation with a halved Kalman gain instead becomes:
\begin{equation}
\hat{\mathbf{X}}^a =  \hat{\mathbf{X}}^b - \frac{1}{2} \tilde{\mathbf{K}} \hat{\mathbf{Y}}^b.
\end{equation}

The sample elements in the EnKF are naturally subject to sampling errors which reduce as $N_e$ increases. %
Localisation \citep{Hamill_et_al2001} is implemented using a straightforward Schur multiplication of \eqref{eq:PHT} and \eqref{eq:HPHT} by an adequate tapering matrix. A compact support approximation to a Gaussian off-diagonal decay is often used for this purpose \citep{Gaspari_Cohn_1999}.

\end{document}